\documentclass[%
 reprint,
 amsmath,amssymb,
 aps,
]{revtex4-2}

\usepackage[english]{babel}
\usepackage{amsmath}
\usepackage{bm}
\usepackage{upgreek}
\usepackage{physics}
\usepackage{amssymb}
\usepackage{color}

\usepackage{graphicx}
\usepackage{dcolumn}
\usepackage{bm}

\makeatletter
\renewcommand{\fnum@figure}{FIG.~\thefigure}
\makeatother

\begin{document}

\preprint{APS/123-QED}

\title{Addition to the dynamic Stark shift of the coherent population trapping resonance}

\author{G.V. Voloshin}
\email{gavriilvsh@gmail.com}
\affiliation{Peter the Great St. Petersburg Polytechnic University (SPbPU), St. Petersburg 195251, Russia}

\author{K.A. Barantsev}
\email{kostmann@yandex.ru}
\affiliation{Peter the Great St. Petersburg Polytechnic University (SPbPU), St. Petersburg 195251, Russia}

\author{A.N. Litvinov}%
\email{andrey.litvinov@mail.ru}
\affiliation{Peter the Great St. Petersburg Polytechnic University (SPbPU), St. Petersburg 195251, Russia}

\begin{abstract}
This paper presents a theoretical study of the light-induced shift of the coherent population trapping resonance. An analytical model is proposed that describes the interaction of two radiation components with an atomic system using a $\Lambda$ scheme and takes into account an additional level of excited state. Both weak and strong coupling regimes with off-resonant transitions are considered. It is shown that, in addition to the conventional dynamic Stark shift, an extra shift arises due to the distortion of the resonance line shape when bichromatic laser radiation interacts with off-resonant atomic transitions. An analytical expression for this additional shift is derived in the weak-coupling limit, and its significant impact on the resonance shape and sensitivity to the intensities of the laser field components is demonstrated. It is found that under strong coupling conditions, the additional shift can deviate substantially from a linear dependence on light intensity, suggesting new opportunities for controlling light shifts in precision atomic devices such as quantum frequency standards.

\end{abstract}

\maketitle

\section{\label{sec:level1} Introduction}

The phenomenon of coherent population trapping (CPT) plays an important role in atomic spectroscopy~\cite{Alzetta1976, Arimondo1976, Gray1978, Agapev1993, Arimondo1996, Wynands1999}, serving as a physical foundation for the development of highly stable quantum frequency standards~\cite{Vanier2005, Zibrov2005, Zibrov2010, Kitching2018, Kobtsev2019, gozzelino2020loaded} and precision sensors~\cite{martinez2023chip, schwindt2004chip, shah2007subpicotesla, mhaskar2012low} due to its narrow resonance linewidth and its ability to effectively suppress Doppler broadening.

CPT occurs when bichromatic laser radiation interacts with a resonant atomic medium. If the optical transitions involved share a common excited state, the two coherent excitation pathways may interfere destructively, causing the atom to become transparent to the incident light. This behavior is resonant with respect to the frequency difference between the two optical components. The spectral width of such a resonance can be made significantly smaller than the natural linewidth of the transition, enabling various practical applications.

The shape of the CPT resonance line strongly depends on the intensity of the excitation field. For practical applications such as frequency standards, one of the main limiting factors is the light shift of the CPT resonance, which occurs both in continuous-wave schemes~\cite{Zhu2000, gerginov2006long, miletic2012ac, chuchelov2018modulation, brazhnikov2024light} and in pulsed Ramsey-type interrogation schemes ~\cite{hemmer1989ac, castagna2009investigations, liu2013ramsey, yano2014theoretical, pati2015computational, abdel2017high, pollock2018ac, voloshin2019effect, voloshin2020effect, voloshin2022line, voloshin2024interference}. In particular, the light shift is a major limitation on the long-term stability of CPT-based atomic clocks~\cite{pollock2022inhomogeneous}. 

A range of methods developed for Ramsey interrogation schemes have been proposed to suppress the light shift in atomic clocks. One group of techniques, known as hyper-Ramsey methods \cite{yudin2010hyper, zanon2015generalized, beloy2018hyper}, employs complex pulse sequences and phase jumps and is primarily used in two-level systems, such as those found in cold-atom clocks. A second class of approaches — auto-balanced Ramsey spectroscopy — relies on modulating the duration of the dark period and implementing a dual-loop feedback mechanism \cite{sanner2018comparison, abdel2018toward, shuker2019ramsey, yudin2018generalized}. These methods can be applied to both two-level systems and CPT schemes. For continuous-wave spectroscopy, which is most commonly used in commercial atomic clocks, techniques for suppressing the light shift are significantly less developed compared to those available for Ramsey spectroscopy. Several approaches to address this issue have been proposed in \cite{shah2006continuous, mcguyer2009simple, gerginov2018two, yudin2020general}.

In most studies addressing the light shift of the CPT resonance, the shift is attributed to the dynamic Stark effect~\cite{bonchbruevich1969, delone1999ac, happer1967effective} and is typically described in the following way. Laser components, resonant with two atomic transitions sharing a common excited state, also interact with off-resonant transitions. For instance, in the D1 line of alkali atoms, the excited state consists of two hyperfine sublevels, of which only one contributes to the resonant transition, while the others interact off-resonantly with the radiation. This interaction results in Stark shifts of the ground state sublevels~\cite{affolderbach2005light, yano2014theoretical, Zhu2000}, which are proportional to the intensities of the excitation components. Since the CPT resonance condition depends on the frequency difference between the two excited transitions, the resonance is shifted by the difference between the two Stark shifts of the ground states.

A limitation of this description is that the influence of off-resonant transitions is not limited to dynamic Stark shifts alone. Excitation of nearby transitions also leads to distortion of the resonance line shape. This distortion results in an additional frequency shift, which in general may be of the same order as the Stark shift and therefore must be taken into account when evaluating the total shift of the observed CPT signal. Despite the significance of this additional contribution, the authors are not aware of any prior studies dedicated to its investigation.

The goal of this work is to provide a theoretical description of the CPT resonance shift caused by distortion of the line shape due to interaction with off-resonant transitions. An analytical formula is derived to estimate this shift based on a model that includes interaction with a nonresonant excited-state sublevel, in the regime of weak coupling. This approximation is valid when the hyperfine splitting of the excited state exceeds the absorption linewidth, a condition often realized in practice. Furthermore, it is shown that in the case of small splitting (strong interaction), the additional shift may lead to a nonlinear dependence of the CPT resonance on the excitation intensity. This result is of particular importance because many methods for compensating light shifts, such as power modulation techniques~\cite{yudin2020general}, are most effective when the shift is linearly dependent on the laser power. 

\section{Mathematical model}
The coherent population trapping resonance arises from the interaction of an atom with a two-frequency field in a $\Lambda$-type configuration. We will consider the resonant case in which the detunings of the two frequency components of the radiation are small compared to the decay rate of the excited state. To account for the light shifts that emerge in such a configuration, it is necessary to consider the interaction of the field's frequency components with atomic transitions that are off-resonant. For a frequency component resonant with the transition $\left| g \right\rangle \to \left| e \right\rangle$, these off-resonant transitions may include either a transition from another ground-state sublevel to the same excited sublevel, $\left| g' \right\rangle \to \left| e \right\rangle$, or a transition from the same ground sublevel to another excited sublevel, $\left| g \right\rangle \to \left| e' \right\rangle$. In this work, we restrict our consideration to the latter case, assuming that the ground-state splitting is much larger than the excited-state splitting and, consequently, that the contribution from the transitions $\left| g \right\rangle \to \left| e' \right\rangle$ dominates the light shift. This assumption is valid, for example, in the case of hyperfine structure in the $D_1$ lines of alkali atoms.

\begin{figure}
\centering
\includegraphics[width=\linewidth]{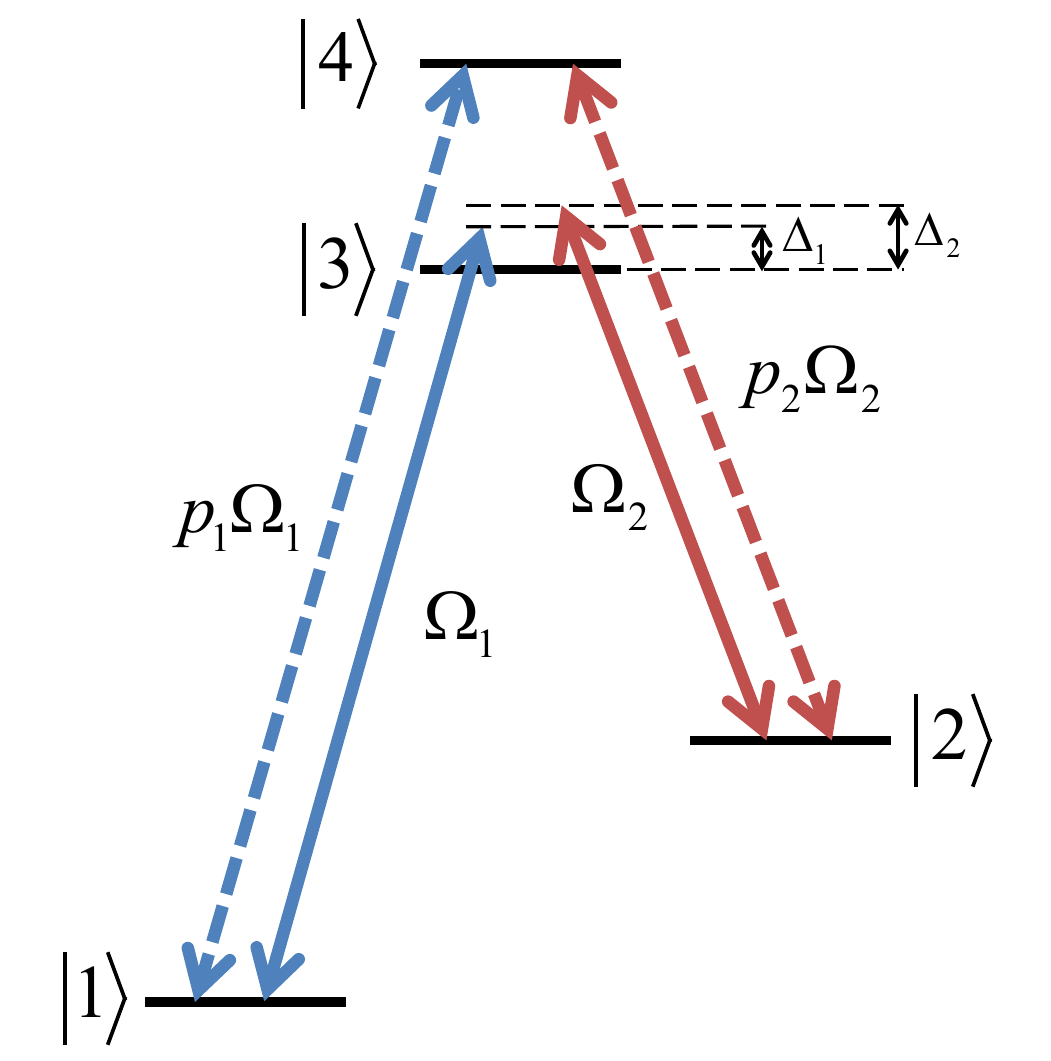}
\caption{Four-level $\Lambda$-type atomic configuration interacting with two-frequency radiation. Solid arrows indicate near-resonant driven transitions. Dashed arrows indicate off-resonant driven transitions. See text for notation.}
\label{level_scheme}
\end{figure}

Accordingly, we adopt a model that describes the light shifts of CPT resonances as a four-level stationary atomic ensemble interacting with classical two-frequency electromagnetic radiation tuned to resonance with transitions to the lower excited-state sublevel (see Fig.~\ref{level_scheme}). This model can correspond to CPT resonance excitation schemes in the $D_1$-lines of alkali atoms using $\sigma^{\pm}\sigma^{\pm}$ polarization configurations, acting on the 0–0 ground-state transitions in the presence of a magnetic field strong enough to resolve the magnetic structure of the CPT resonance. The latter assumption implies that the influence of neighboring Zeeman transitions is negligible, since the applied magnetic field spectrally isolates the selected 0–0 transitions.

Within the semiclassical approach, the two-frequency electromagnetic field is described by a classical electric field vector $\mathbf{E}$, which depends on time at the atomic position as follows:

\begin{equation}
\mathbf{E}(t)= \frac{1}{2} {\mathbf{E}_{1}}{{e}^{-i{{\omega }_{1}}t}}+\frac{1}{2}{\mathbf{E}_{2}}{{e}^{-i{{\omega }_{2}}t}}+\text{c.c.},
\end{equation}
where $\mathbf{E}_{g}$ and $\omega_{g}$ are the complex amplitude vectors and angular frequencies of the two field components, respectively ($g=1,2$), and c.c. denotes the complex conjugate of the preceding terms.

The quantum state of atoms interacting with their environment is mixed and is therefore described by a density matrix $\hat{\rho}$ satisfying the quantum master equation:
\begin{equation}
\frac{\partial }{\partial t}\hat{\rho }=-\frac{i}{\hbar} \left[ \hat{H},\hat{\rho } \right]+\hat{\hat{R}}\left\{ {\hat{\rho }} \right\},
\end{equation}
where $\hat{H}$ is the atomic Hamiltonian and $\hat{\hat{R}}\left\{ {\hat{\rho }} \right\}$ is a superoperator accounting for relaxation of the excited states. The Hamiltonian $\hat{H}$ is written as the sum
\begin{equation}
\hat{H}={{\hat{H}}_{0}}+\hat{V},
\end{equation}
with the free atomic Hamiltonian
\[{{\hat{H}}_{0}}=\sum\limits_{n=1}^{4}{{{\varepsilon }_{n}}\left| n \right\rangle \left\langle  n \right|}\]
and the interaction Hamiltonian in the dipole approximation
\begin{equation*}
\begin{aligned}
\hat{V}=&-\hat{\mathbf{d}}\mathbf{E}=-\hbar {{e}^{-i{{\omega }_{1}}t}}\left( {{\Omega }_{1}}\left| 1 \right\rangle \left\langle  3 \right|+{{p}_{1}}{{\Omega }_{1}}\left| 1 \right\rangle \left\langle  4 \right| \right)\\
&-\hbar {{e}^{-i{{\omega }_{2}}t}}\left( {{\Omega }_{2}}\left| 2 \right\rangle \left\langle  3 \right|+{{p}_{2}}{{\Omega }_{2}}\left| 2 \right\rangle \left\langle  4 \right| \right)+\text{H.c.}
\end{aligned}
\end{equation*}
Here, $\hat{\mathbf{d}}$ is the dipole moment operator; $\Omega_{g} = d_{g3} E_{g} / (2\hbar)$ are the half-Rabi frequencies for the transitions $\left| g \right\rangle \to \left| 3 \right\rangle$; $p_{g} = d_{g4} / d_{g3}$ are the ratios of dipole matrix elements for the transitions $\left| g \right\rangle \to \left| 4 \right\rangle$ and $\left| g \right\rangle \to \left| 3 \right\rangle$; $d_{ge}$ are the projections of the dipole matrix elements along the polarization directions of $\mathbf{E}_{g}$ ($e=3,4$). The H.c. term denotes the Hermitian conjugate of the preceding terms. We neglect cross terms $\sim \Omega_{g} |g'\rangle\langle e|$ under the assumption that the frequencies $\omega_{g}$ are significantly detuned from the $\left| g' \right\rangle \to \left| e \right\rangle$ transitions (much more than from $\left| g \right\rangle \to \left| e' \right\rangle$). We also assume that the electric-dipole transitions $\left| g \right\rangle \to \left| g' \right\rangle$ and $\left| e \right\rangle \to \left| e' \right\rangle$ are forbidden.

The matrix elements of the relaxation superoperator $\hat{\hat{R}}$ are written assuming exponential decay of the excited states and atomic coherences:
\begin{equation}
\begin{aligned}
  {{R}_{ee}}\left\{ {\hat{\rho }} \right\}&=-{{\gamma }_{e}}{{\rho }_{ee}}, \\ 
 {{R}_{gg}}\left\{ {\hat{\rho }} \right\}&={{\gamma }_{3g}}{{\rho }_{33}}+{{\gamma }_{4g}}{{\rho }_{44}}, \\ 
 {{R}_{nm}}\left\{ {\hat{\rho }} \right\}&=-{{\Gamma }_{nm}}{{\rho }_{nm}},\text{ }n\ne m,
\end{aligned}
\end{equation}
where $\rho_{nm} = \langle n | \hat{\rho} | m \rangle$ ($n,m = 1,\dots,4$), and $\gamma_{ee}$, $\gamma_{eg}$, $\Gamma_{nm}$ are the decay rates of excited states, the population transfer rates to ground states, and the decoherence rates, respectively. Conservation of normalization, $\sum_{n} \dot{\rho}_{nn} = 0$, implies $\gamma_{e1} + \gamma_{e2} = \gamma_{e}$. The decay rates of optical coherences $\Gamma_{ge}$ include both spontaneous emission and atom-atom collisions, such that $\Gamma_{ge} \ge \gamma_{e}/2$. We assume that $\Gamma_{ge}$ is larger than the Doppler width: $\Gamma_{ge} > k_g v_T$, where $v_T$ is the thermal velocity of atoms, $k_g$ - wave number of the field $\mathbf{E}_g$. This justifies using a model of stationary atoms.

By expressing Eq. (2) in components and factoring out rapidly oscillating terms:
\begin{equation}
\begin{aligned}
  & {{\rho }_{ge}}=\rho _{eg}^{*}={{e}^{i{{\omega }_{g}}t}}{{{\tilde{\rho }}}_{ge}}, \\ 
 & {{\rho }_{12}}=\rho _{21}^{*}={{e}^{i\left( {{\omega }_{1}}-{{\omega }_{2}} \right)t}}{{{\tilde{\rho }}}_{12}}. \\ 
\end{aligned}
\end{equation}

Substituting Eqs. (3)–(5) into (2), we obtain
\begin{equation}
\begin{aligned}
{{\dot{\rho }}_{11}}=&-i\left( {{\Omega }_{1}}{{{\tilde{\rho }}}_{13}}-\Omega _{1}^{*}{{{\tilde{\rho }}}_{31}}+{{p}_{1}}{{\Omega }_{1}}{{{\tilde{\rho }}}_{14}}-p_{1}^{*}\Omega _{1}^{*}{{{\tilde{\rho }}}_{41}} \right) \\
& +{{\gamma }_{31}}{{\rho }_{33}}+{{\gamma }_{41}}{{\rho }_{44}}, \\
{{\dot{\rho }}_{22}}=&-i\left( {{\Omega }_{2}}{{{\tilde{\rho }}}_{23}}-\Omega _{2}^{*}{{{\tilde{\rho }}}_{32}}+{{p}_{2}}{{\Omega }_{2}}{{{\tilde{\rho }}}_{24}}-p_{2}^{*}\Omega _{2}^{*}{{{\tilde{\rho }}}_{42}} \right) \\
& +{{\gamma }_{32}}{{\rho }_{33}}+{{\gamma }_{42}}{{\rho }_{44}}, \\
{{\dot{\rho }}_{33}}=&i{{\Omega }_{1}}{{\tilde{\rho }}_{13}}-i\Omega _{1}^{*}{{\tilde{\rho }}_{31}}+i{{\Omega }_{2}}{{\tilde{\rho }}_{23}}-i\Omega _{2}^{*}{{\tilde{\rho }}_{32}}-{{\gamma }_{3}}{{\rho }_{33}}, \\
{{\dot{\rho }}_{44}}=&i{{p}_{1}}\Omega _{1}^{{}}{{\tilde{\rho }}_{14}}-ip_{1}^{*}\Omega _{1}^{*}{{\tilde{\rho }}_{41}}+i{{p}_{2}}{{\Omega }_{2}}{{\tilde{\rho }}_{24}}-ip_{2}^{*}\Omega _{2}^{*}{{\tilde{\rho }}_{42}} \\
& -{{\gamma }_{4}}{{\rho }_{44}},\\
{{\dot{\tilde{\rho }}}_{13}}=& ip_{1}^{*}\Omega _{1}^{*}{{\rho }_{43}}+i\Omega _{1}^{*}{{\rho }_{33}}-i\Omega _{2}^{*}{{\rho }_{12}}-i\Omega _{1}^{*}{{\rho }_{11}} \\
& -\left[ i{{\Delta }_{1}}+{{\Gamma }_{13}} \right]{{\rho }_{13}},\\
{{\dot{\tilde{\rho }}}_{14}}=& i\Omega _{1}^{*}{{\rho }_{34}}+ip_{1}^{*}\Omega _{1}^{*}{{\rho }_{44}}-ip_{2}^{*}\Omega _{2}^{*}{{\rho }_{12}}-ip_{1}^{*}\Omega _{1}^{*}{{\rho }_{11}} \\
& -\left[ i\left( {{\Delta }_{1}}-{{\omega }_{34}} \right)+{{\Gamma }_{14}} \right]{{\rho }_{14}},\\
{{\dot{\tilde{\rho }}}_{23}}=& ip_{2}^{*}\Omega _{2}^{*}{{\rho }_{43}}+i\Omega _{2}^{*}{{\rho }_{33}}-i\Omega _{1}^{*}{{\rho }_{21}}-i\Omega _{2}^{*}{{\rho }_{22}} \\
& -\left[ i{{\Delta }_{2}}+{{\Gamma }_{23}} \right]{{\rho }_{23}},\\
{{\dot{\tilde{\rho }}}_{24}}=& i\Omega _{2}^{*}{{\rho }_{34}}+ip_{2}^{*}\Omega _{2}^{*}{{\rho }_{44}}-ip_{1}^{*}\Omega _{1}^{*}{{\rho }_{21}}-ip_{2}^{*}\Omega _{2}^{*}{{\rho }_{22}} \\
& -\left[ i\left( {{\Delta }_{2}}-{{\omega }_{34}} \right)+{{\Gamma }_{24}} \right]{{\rho }_{24}}, \\
{{\dot{\rho }}_{34}}=&i{{\Omega }_{1}}{{\tilde{\rho }}_{14}}-i{{p}_{1}}\Omega _{1}^{*}{{\tilde{\rho }}_{31}}+i{{\Omega }_{2}}{{\tilde{\rho }}_{24}}-i{{p}_{2}}\Omega _{2}^{*}{{\tilde{\rho }}_{32}} \\
& -\left[ -i{{\omega }_{34}}+{{\Gamma }_{34}} \right]{{\rho }_{34}}, \\
{{\dot{\tilde{\rho }}}_{12}}=&i\left( \Omega _{1}^{*}{{{\tilde{\rho }}}_{32}}-{{\Omega }_{2}}{{{\tilde{\rho }}}_{13}}+p_{1}^{*}\Omega _{1}^{*}{{{\tilde{\rho }}}_{42}}-{{p}_{2}}{{\Omega }_{2}}{{{\tilde{\rho }}}_{14}} \right) \\
& -\left[ i\delta +{{\Gamma }_{12}} \right]{{\tilde{\rho }}_{12}}.\\
\end{aligned}
\end{equation}
where $\omega_{34}$ is the frequency of the transition $\left| 3 \right\rangle \to \left| 4 \right\rangle$, $\Delta_{g}$ are the detunings of $\omega_{g}$ from the transitions $\left| g \right\rangle \to \left| 3 \right\rangle$ (see Fig.~\ref{level_scheme}), and $\delta = \Delta_{1} - \Delta_{2}$ is the two-photon detuning. Terms oscillating near $e^{\pm 2i\omega_{g}t}$ are neglected under the rotating wave approximation (RWA) \cite{scully1997quantum}.

We further restrict ourselves to the uniform relaxation model \cite{happer1972optical}:

\begin{equation}
\begin{aligned}
  & {{\gamma }_{3}}={{\gamma }_{4}}\equiv \gamma , \\
  & {{\gamma }_{13}}={{\gamma }_{23}}={{\gamma }_{14}}={{\gamma }_{24}} = \gamma/2 , \\
  & {{\Gamma }_{13}}={{\Gamma }_{23}}={{\Gamma }_{14}}={{\Gamma }_{24}}\equiv \Gamma .
\end{aligned}
\end{equation}

Additionally, we apply the adiabatic approximation ($\Omega_{g} \ll \Gamma$), which allows us to consider the populations of the excited states and the coherence $\rho_{34}$ as small compared to the populations of the ground states: $\rho_{ee}, |\rho_{34}| \ll \rho_{gg}$.

\section{Results and discussion}
The presence of a fourth level in the excitation scheme has a smaller effect on the excitation process the greater the splitting $\omega_{34}$ of the excited state compared to the decay rate $\Gamma$, or the smaller the values of $p_g$. In both cases, the interaction of radiation with the fourth level can be considered weak compared to its interaction with the third level. In this section, we analyze the CPT resonance shifts for both weak and strong interactions. The weak interaction case is of particular interest since it is more common in real excitation schemes and allows for a simple analytical expression for the CPT resonance shift.

\subsection{Case of weak interaction with the fourth level}

For definiteness, we define the weak interaction approximation as the case when $\omega_{34} \gg \Gamma$. Note, however, that the results remain unchanged if this condition is replaced with $|p_g| \ll 1$.

In the weak interaction approximation, the solution to system (6) can be sought as a power series expansion in the small parameter $\Gamma/\omega_{34}$, with the zeroth-order term corresponding to the solution of an analogous three-level $\Lambda$ system without the fourth level. We obtain the solution for the low-frequency coherence $\tilde{\rho}_{12}$ and the ground-state populations $\rho_{gg}$ in the region of small detunings $\Delta_g \ll \Gamma$ under stationary conditions, retaining all first-order terms in $\Gamma/\omega_{34}$ (see Appendix A):

\begin{equation}
    {{\tilde{\rho }}_{12}}=\tilde{\rho }_{12}^{(0)}+ip_{1}^{*}{{p}_{2}}G\frac{{{\left| {{\Omega }_{2}} \right|}^{2}}-{{\left| {{\Omega }_{1}} \right|}^{2}}}{{{\left| {{\Omega }_{1}} \right|}^{2}}+{{\left| {{\Omega }_{2}} \right|}^{2}}}\tilde{\rho }_{12}^{(0)},
\end{equation}

\begin{equation}
    {{\rho }_{gg}}=\frac{{{\left| {{\Omega }_{g'}} \right|}^{2}}}{{{\left| {{\Omega }_{1}} \right|}^{2}}+{{\left| {{\Omega }_{2}} \right|}^{2}}}+{{\left( -1 \right)}^{g+1}}\frac{2p_{1}^{*}{{p}_{2}}G\Gamma \left( \delta -{{\delta }_{AC}} \right)}{{{\left| {{\Omega }_{1}} \right|}^{2}}+{{\left| {{\Omega }_{2}} \right|}^{2}}}{{\left| \tilde{\rho }_{12}^{(0)} \right|}^{2}},
\end{equation}
where \[\tilde{\rho}_{12}^{(0)} = -\frac{\Omega_1^* \Omega_2}{\Gamma [\gamma_D + i(\delta - \delta_{AC})]}\] is the stationary low-frequency coherence for the analogous three-level $\Lambda$-system, shifted along the $\delta$ axis by
\[{{\delta }_{AC}}=\frac{G}{\Gamma }\left( {{\left| {{p}_{1}}{{\Omega }_{1}} \right|}^{2}}-{{\left| {{p}_{2}}{{\Omega }_{2}} \right|}^{2}} \right);\]
\[G = \frac{\Gamma \omega_{34}}{\Gamma^2 + \omega_{34}^2}\] is a dimensionless function of $\omega_{34}$, which has a dispersive N-shaped contour; \[\gamma_D = \frac{|\Omega_1|^2 + |\Omega_2|^2}{\Gamma} + \Gamma_{12}\] is the CPT resonance linewidth; ${g}'\ne g$. In (8) and (9), we account for the fact that $G$ is a first-order small quantity in the parameter $\Gamma/\omega_{34}$. Note that $\delta_{AC}$ is the shift of level $\left| 3 \right\rangle$ due to the bichromatic radiation interacting with transitions $|g\rangle \to |4\rangle$, given by the well-known formula for the dynamic Stark effect \cite{affolderbach2005light, yano2014theoretical, Zhu2000}.

\begin{figure}
\centering
\includegraphics[width=\linewidth]{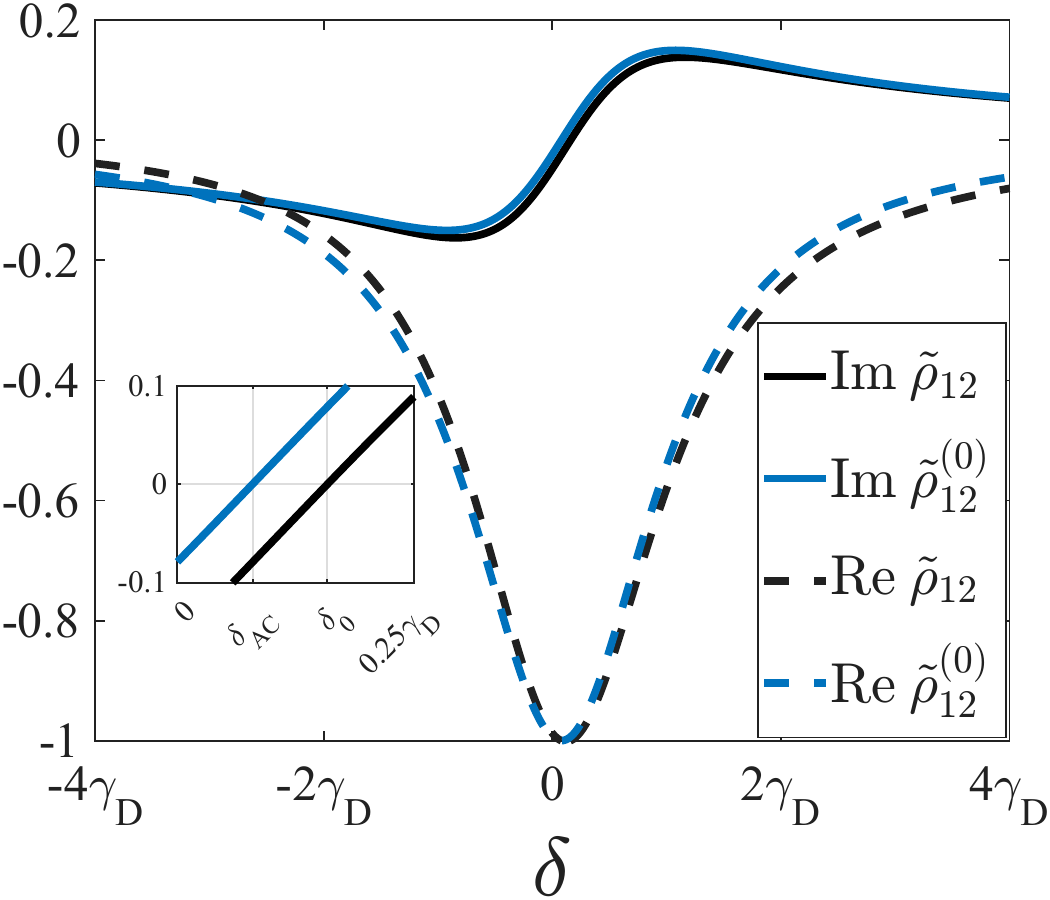}
\caption{Dependence of the real and imaginary parts of $\tilde{\rho}_{12}$ and $\tilde{\rho}_{12}^{(0)}$ in relative units on the two-photon detuning $\delta$ for $\omega_{34} = 10\Gamma$, $\Omega_1 = 3\Omega_2$, $|\Omega_1|^2 + |\Omega_2|^2 = 10^{-4}\Gamma^2$, $\Gamma_{12} = 0$, $p_1 = 1$, $p_2 = -1$.}
\label{rho12_delta}
\end{figure}

Thus, from (8), we see that the interaction with the fourth sublevel, in first-order approximation, not only shifts the CPT resonance by $\delta_{AC}$ but also distorts its shape, as described by the second term in Eq. (8). This distortion results in an additional shift. We illustrate this shift by analyzing the in-phase and quadrature components of the low-frequency coherence $\tilde{\rho}_{12}$ versus two-photon detuning (see Fig.~\ref{rho12_delta}). As shown in Fig.~\ref{rho12_delta}, the symmetry of $\tilde{\rho}_{12}(\delta)$ with respect to the point $\delta_{AC}$ disappears when interaction with the fourth level is taken into account. Consequently, the zero point $\delta_0$ of the imaginary part of $\tilde{\rho}_{12}$ does not coincide with $\delta_{AC}$.

Let us find the value of ${{\delta }_{0}}$ from the condition
\begin{equation}
    {{\left. \operatorname{Im}{{{\tilde{\rho }}}_{12}} \right|}_{\delta ={{\delta }_{0}}}}=0.
\end{equation}
In the case of real Rabi frequencies ${{\Omega }_{g}}$ and ratios ${{p}_{g}}$, we obtain:
\begin{equation}
    {{\delta }_{0}}={{\delta }_{AC}}+{{\delta }_{D}},
\end{equation}
where
\begin{equation}
    {{\delta }_{D}}={{p}_{1}}{{p}_{2}}G{{\gamma }_{D}}\frac{\Omega _{2}^{2}-\Omega _{1}^{2}}{\Omega _{1}^{2}+\Omega _{2}^{2}}
\end{equation}
is the additional shift caused by the distortion of the CPT resonance line shape due to the interaction of the field with off-resonant transitions from the ground state to the fourth level. Here, the modulus signs on ${{\Omega }_{g}}$ are omitted since the Rabi frequencies are assumed to be real.

\begin{figure}
\centering
\includegraphics[width=\linewidth]{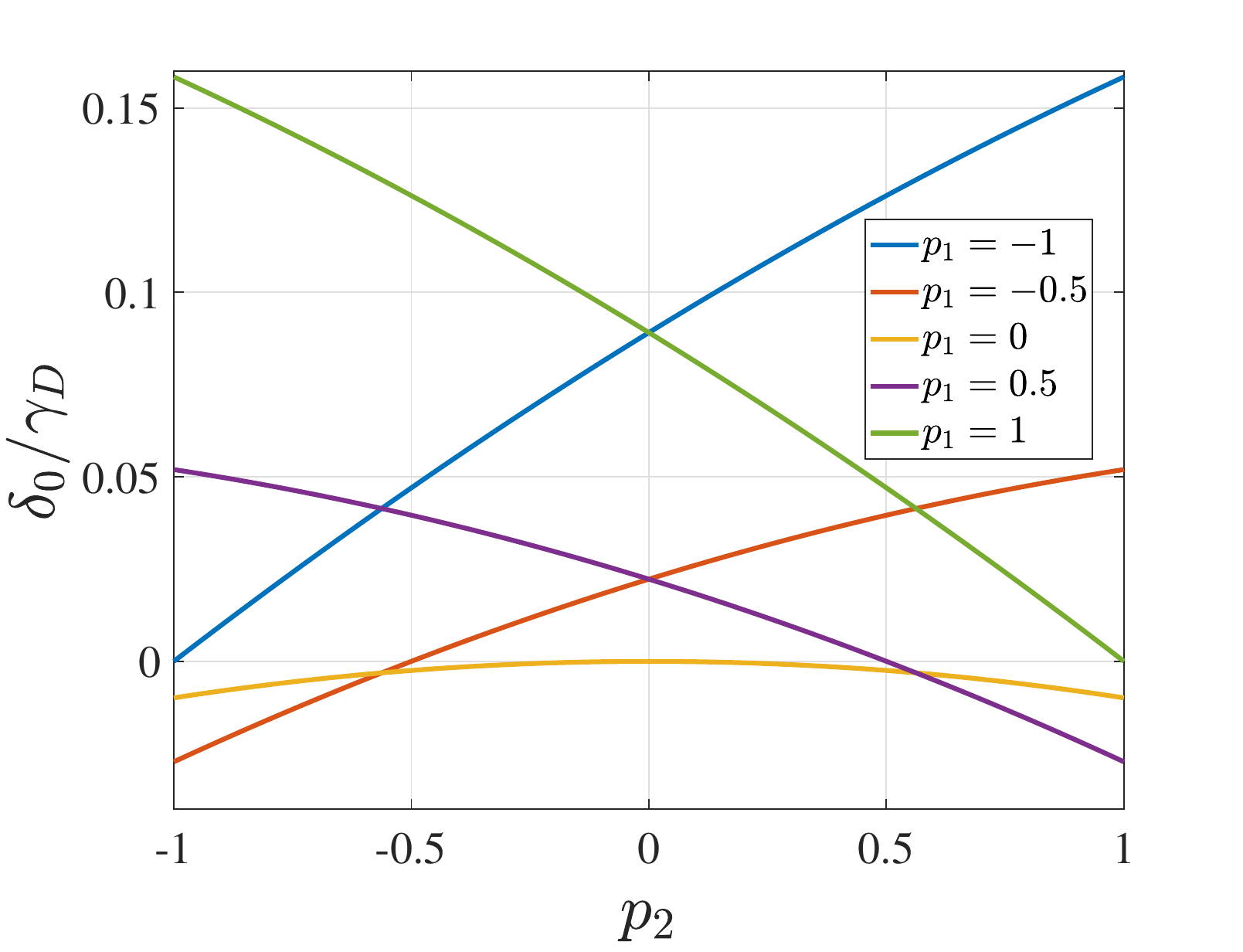}
\caption{Dependence of the total shift $\delta_0$ on the parameter $p_2$ in units of $\gamma_D$ for different $p_1$. The remaining parameters are the same as in the caption to Fig. \ref{rho12_delta}}
\label{delta0_p1_p2}
\end{figure}

The dependence of the obtained total shift $\delta_0$ on $p_1$ and $p_2$, described by Eq.~(11), is shown in Fig. 3. The result of adding an additional shift $\delta_D$ is a shift in the extremum of the parabolic dependence of the $\delta_0$ on $p_g$ from the zero value

Let us note some features of the obtained shift ${{\delta }_{0}}$ as compared to the standard Stark shift ${{\delta }_{AC}}$. First, the additional shift ${{\delta }_{D}}$ is of the same order of magnitude as ${{\delta }_{AC}}$ and therefore must be taken into account when evaluating the CPT resonance shift. Second, at sufficiently low radiation intensities, when the resonance width ${{\gamma }_{D}}$ is mainly determined by the value of ${{\Gamma }_{12}}$, the additional shift ${{\delta }_{D}}$ is proportional to the relative difference in the intensities of the two frequency components of the field, unlike the shift ${{\delta }_{AC}}$, which depends on their absolute difference. Third, the additional shift ${{\delta }_{D}}$, in contrast to ${{\delta }_{AC}}$, depends on the signs of the quantities ${{p}_{g}}$. For example, in the case ${{p}_{1}}={{p}_{2}}$ and ${{\Gamma }_{12}}=0$, the shifts ${{\delta }_{AC}}$ and ${{\delta }_{D}}$ cancel each other and ${{\delta }_{0}}=0$, which is not the case for ${{p}_{1}}=-{{p}_{2}}$ (see Fig. \ref{delta0_p1_p2}). This last remark remains valid beyond the weak interaction approximation and can be explained by the fact that the considered excitation scheme forms a closed loop $\left| 1 \right\rangle \leftrightarrow \left| 3 \right\rangle \leftrightarrow \left| 2 \right\rangle \leftrightarrow \left| 4 \right\rangle \leftrightarrow \left| 1 \right\rangle$, which, as is well known \cite{matisov1992sensitivity, kazakov2006dark}, leads to resonance sensitivity to the excitation phase. All this allows us to conclude that the additional contribution ${{\delta }_{D}}$ to the CPT resonance shift is significant.

However, the value of ${{\tilde{\rho }}_{12}}$ is not directly observable. In practice, the absorption signal, determined by the imaginary parts of the complex dielectric susceptibilities ${{\chi }_{g}}$, is often used as the spectral signal. Taking into account that both components of the field interact with two transitions, we define the susceptibilities, omitting dimensional constants, as follows:
\begin{equation}
    {{\chi }_{g}}=\frac{1}{{{\Omega }_{g}}}\left( {{{\tilde{\rho }}}_{3g}}+{{p}_{g}}{{{\tilde{\rho }}}_{4g}} \right).
\end{equation}
Here, we use the fact that the optical coherences ${{\tilde{\rho }}_{eg}}$ are proportional to the complex polarizations induced by the field components ${{\Omega }_{g}}$ on the transitions $\left| g \right\rangle \to \left| e \right\rangle$ \cite{scully1997quantum}.

In CPT studies, the fluorescence signal is also frequently used as a spectral observable, being proportional to the dependence of the excited-state population on the two-photon detuning. We consider the total excited-state population ${{\rho }_{exc}}={{\rho }_{33}}+{{\rho }_{44}}$. In the stationary case, from Eq. (6) we find:
\begin{eqnarray}
    &&{{\rho }_{\text{exc}}}=\nonumber\\
    &&-\frac{2}{\gamma }\operatorname{Im}\left\{ {{\Omega }_{1}}{{{\tilde{\rho }}}_{13}}+{{\Omega }_{2}}{{{\tilde{\rho }}}_{23}}+{{p}_{1}}{{\Omega }_{1}}{{{\tilde{\rho }}}_{14}}+{{p}_{2}}{{\Omega }_{2}}{{{\tilde{\rho }}}_{24}} \right\}=\nonumber\\
    &&-\frac{2}{\gamma }\left( {{\left| {{\Omega }_{1}} \right|}^{2}}{{\chi }''_{1}}+{{\left| {{\Omega }_{2}} \right|}^{2}}{{\chi }''_{2}} \right),
\end{eqnarray}
where ${\chi }''_{g}=\operatorname{Im}{{\chi }_{g}}$.

Let us determine the resonance shifts of the absorption signal and ${{\rho }_{\text{exc}}}$ in the case of real ${{\Omega }_{g}}$ and ${{p}_{g}}$. We express the optical coherences from Eq. (6) in the region of small detunings ${{\Delta }_{g}}\ll \Gamma $, neglecting second-order small quantities in the parameter $\Gamma /{{\omega }_{34}}$:
\begin{equation*}
    \begin{aligned}
{{\rho }_{g3}}&=-\frac{i}{\Gamma }\left( {{\Omega }_{g}}{{\rho }_{gg}}+{{\Omega }_{{{g}'}}}{{{\tilde{\rho }}}_{g{g}'}} \right), \\
{{\rho }_{g4}}&=\frac{G}{\Gamma }\left( {{p}_{g}}{{\Omega }_{g}}{{\rho }_{gg}}+{{p}_{{{g}'}}}{{\Omega }_{{{g}'}}}{{{\tilde{\rho }}}_{g{g}'}} \right).
    \end{aligned}
\end{equation*}

Substituting these expressions into Eq. (13), we obtain:
\begin{equation}
{{\chi }_{g}}=\frac{i}{\Gamma {{\Omega }_{g}}}\left[ {{\Omega }_{g}}{{\rho }_{gg}}\left( 1-iGp_{g}^{2} \right)+{{\Omega }_{{{g}'}}}{{{\tilde{\rho }}}_{{g}'g}}\left( 1-iG{{p}_{1}}{{p}_{2}} \right) \right].
\end{equation}
The imaginary parts of expression (15) are, up to a sign, proportional to the absorption coefficients of the two frequency components of the radiation. From (15), we obtain:
\[
{{\chi }''_{g}}=-\frac{\Omega _{{{g}'}}^{2}}{\Gamma {{\gamma }_{D}}}\frac{\gamma _{D}^{2}+2{{\delta }_{D}}\left( \delta -{{\delta }_{AC}} \right)}{\gamma _{D}^{2}+{{\left( \delta -{{\delta }_{AC}} \right)}^{2}}}+\frac{\Omega _{{{g}'}}^{2}}{\Omega _{1}^{2}+\Omega _{2}^{2}},
\]
from which it follows that the extrema of the absorption coefficients, in the first-order expansion in the small parameter $G$, are achieved at the same point $\delta ={{\delta }_{AC}}+{{\delta }_{D}}$ as the zero of the imaginary part of the low-frequency coherence ${{\delta }_{0}}$. According to (14), the extremum of the ${{\rho }_{\text{exc}}}$ signal is located at the same point.

Thus, expression (11) defines the shifts of the observable CPT resonances in the case of weak interaction of the field with the fourth level. We emphasize that this is valid only within the uniform decay model (7).

It should be noted that although the results presented here were obtained within the model of stationary atoms, they can be straightforwardly generalized to the case of atomic motion in a buffer gas medium, provided that boundary and diffusion effects are neglected. In this case, it is sufficient to account for the Doppler frequency shift, which affects only the velocity distribution of the optical coherences $\tilde{\rho}_{ge}$, under the assumption that the laser field has a negligible effect on the equilibrium Maxwellian distribution of the low-frequency coherence $\tilde{\rho}_{12}$ \cite{voloshin2020effect}. Averaging over the atomic velocities then results in Doppler broadening of the corresponding spectral profiles. In the present model, this only leads to a redefinition of the parameters $\Gamma$ and $G$.

\subsection{Case of strong interaction with the fourth level}

In the case of small ${{\omega }_{34}}$, when the parameter $\Gamma /{{\omega }_{34}}$ is large, the solutions (8)–(9) lose their applicability, and to calculate the shift it is necessary to find an exact solution of the system (6). Hereafter, we restrict ourselves to the analysis of the case ${{\Gamma }_{12}}=0$, since a nonzero value of ${{\Gamma }_{12}}$ weakly affects the shape of the spectral signal at not too low intensities (${{\Omega }_{g}}\gg \sqrt{{{\Gamma }_{12}}\Gamma }$). By finding the optical coherences from (6) and substituting them into (13), we obtain the imaginary parts of the susceptibilities ${\chi }''_{g}$, whose dependence on the two-photon detuning $\delta$ has the form
\begin{equation}
    {{\chi }''_{g}}=\Gamma \Omega _{{{g}'}}^{2}\frac{\sum\limits_{n=0}^{7}{{{A}_{n}}{{\delta }^{n}}}}{\sum\limits_{n=0}^{9}{{{B}_{n}}{{\delta }^{n}}}},
\end{equation}
where ${{A}_{n}},{{B}_{n}}$ are real coefficients depending on the model parameters (see Appendix B).

\begin{figure*}
\centering
\includegraphics[width=90mm]{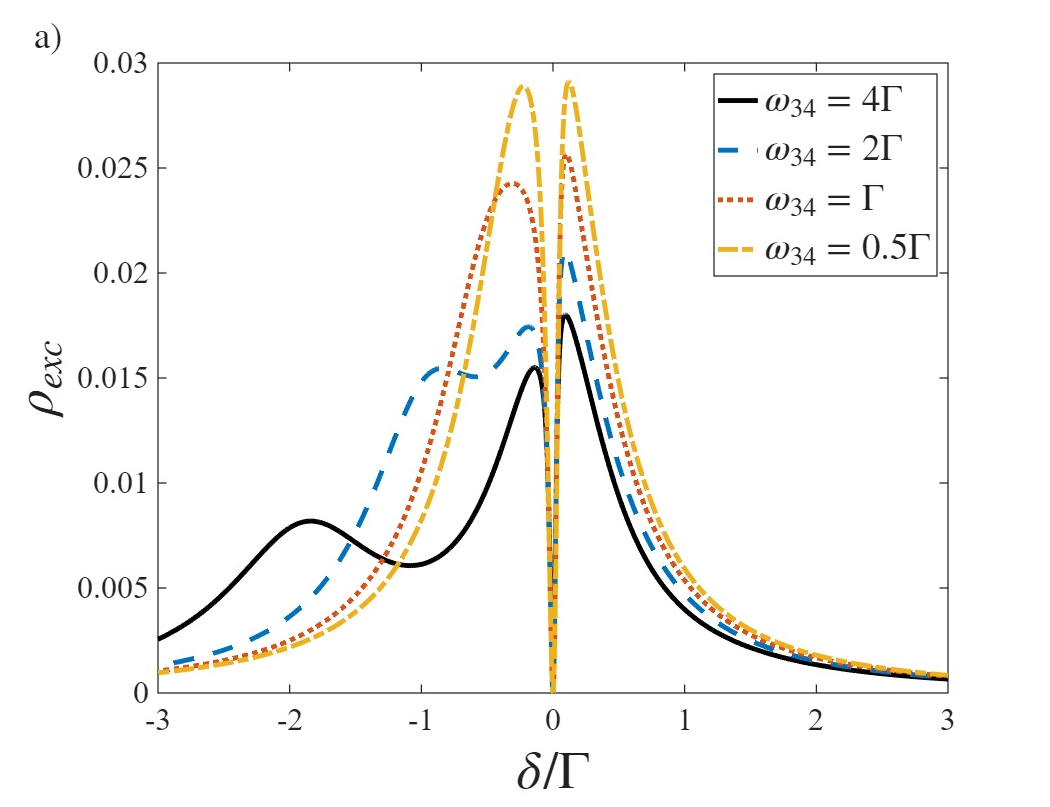}\includegraphics[width=90mm]{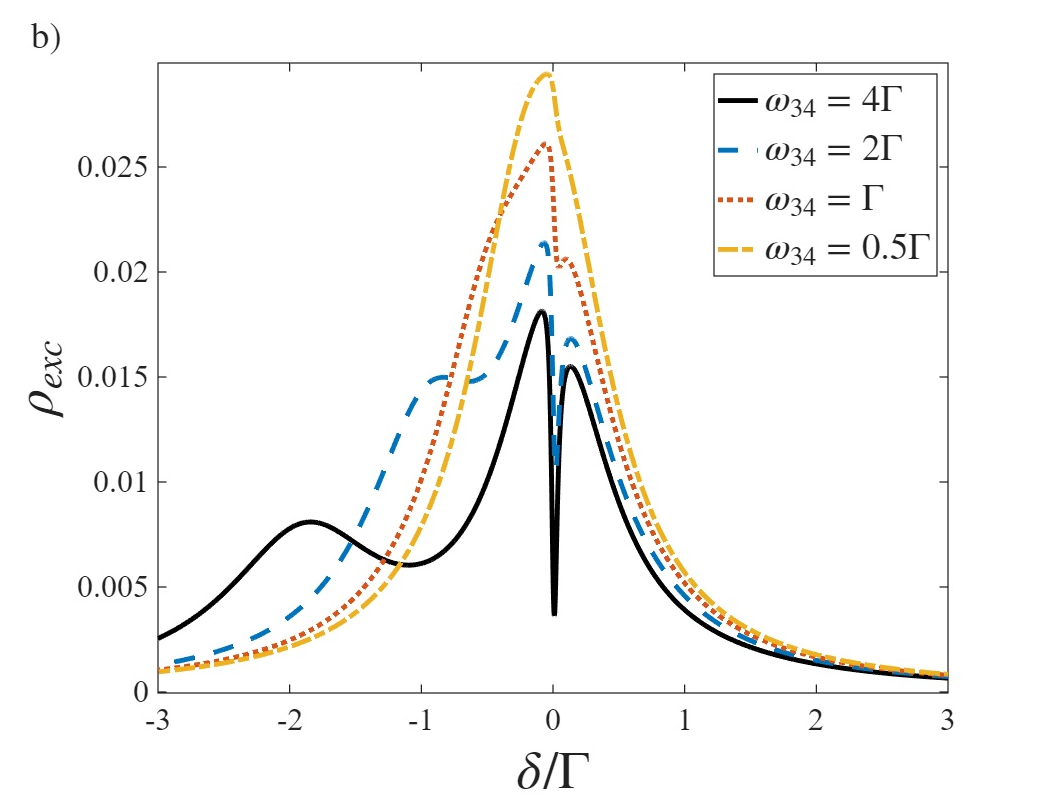}
\caption{Dependence of the CPT resonance shapes observed in the signal ${{\rho }_{\text{exc}}}$ on the excited state splitting ${{\omega }_{34}}$ for cases (a) ${{p}_{1}}={{p}_{2}}=1$ and (b) ${{p}_{1}}=-1,{{p}_{2}}=1$ at $\Omega _{1}^{2}=10\Omega _{2}^{2}$, $\Omega _{1}^{2}+\Omega _{2}^{2}=0.1{{\Gamma }^{2}}$.}
\label{rho_exc}
\end{figure*}

Substituting these into (14), we find the dependence ${{\rho }_{\text{exc}}}(\delta)$. We analyze this dependence as ${{\omega }_{34}}$ decreases relative to $\Gamma$ (Fig.~\ref{rho_exc}) for two cases: ${{p}_{1}}={{p}_{2}}=1$ and ${{p}_{1}}=-1,{{p}_{2}}=1$. These cases correspond to the excitation of resonances in the D1 lines of alkali atoms via polarizations ${{\sigma }^{\pm }}{{\sigma }^{\pm }}$ and ${{\sigma }^{\pm }}{{\sigma }^{\mp }}$, respectively (see, e.g., Ref. \cite{steck2016rubidium}).

Fig.~\ref{rho_exc}b shows that in the case where ${{p}_{1}}$ and ${{p}_{2}}$ have opposite signs, the CPT resonance disappears as ${{\omega }_{34}}$ decreases. This is due to the fact that in the emerging closed excitation loop $\left| 1 \right\rangle \leftrightarrow \left| 3 \right\rangle \leftrightarrow \left| 2 \right\rangle \leftrightarrow \left| 4 \right\rangle \leftrightarrow \left| 1 \right\rangle$, the corresponding optical coherences interfere destructively \cite{matisov1992sensitivity, kazakov2006dark}, unlike the case ${{p}_{1}}={{p}_{2}}=1$, where the CPT resonance persists [Fig.~\ref{rho_exc}(a)]. These cases are realized in CPT resonance excitation schemes in the D1 lines on transitions with ground-state sublevels having total angular momentum projection $m=0$ via polarization configurations $\text{lin} \! \perp \! \text{lin}$ \cite{zanon2005high} and $\text{lin} \! \parallel \! \text{lin}$ \cite{taichenachev2005unique}, respectively.

Let us now discuss the CPT resonance shift in the case of small ${{\omega }_{34}}$. Finding the extremum point of the function ${{\chi }''_{g}}(\delta)$ leads to the equation
\begin{equation}
    {{\left. \frac{\partial }{\partial \delta }{{\chi }''_{g}} \right|}_{\delta ={{\delta }_{0}}}}=0,
\end{equation}
which reduces to a 15th-degree polynomial equation in the unknown ${{\delta }_{0}}$. The solution of equation (17) must be found numerically.

\begin{figure}
\centering
\includegraphics[width=\linewidth]{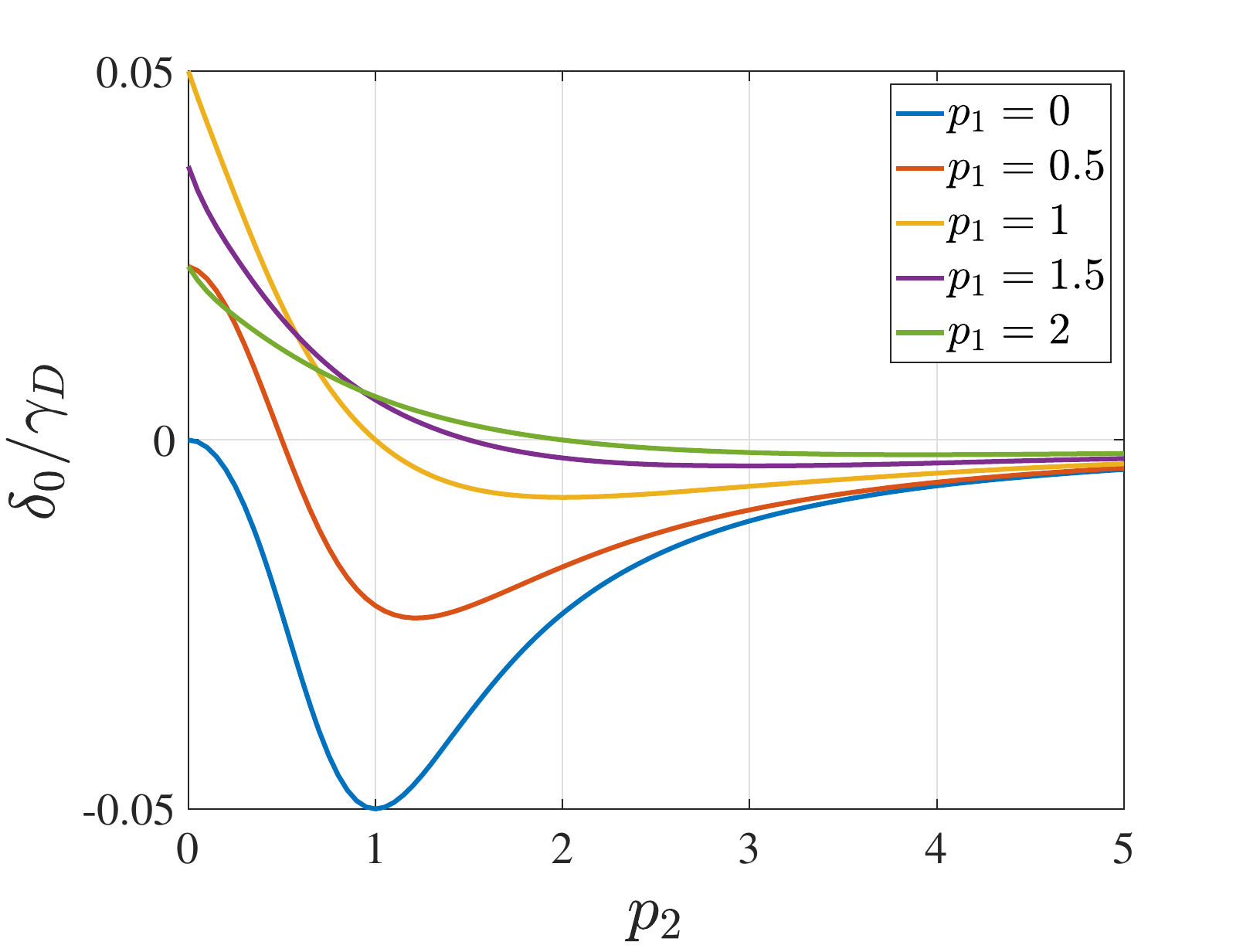}
\caption{Dependence of the shift $\delta_0$ on $p_2$ in units of $\gamma_D$ for different $p_1$ in the strong coupling regime with the fourth level ($\omega_{34}=0.5\Gamma$) for case $\Omega _{1}=\Omega _{2}$, $\Omega _{1}^{2}+\Omega _{2}^{2}=10^{-4}{{\Gamma }^{2}}$.}.
\label{delta0_p1_p2_strong}
\end{figure}

In accordance with the discussion of Fig. 4, the case where $p_1$ and $p_2$ have the same sign is of particular interest. Fig. \ref{delta0_p1_p2_strong} shows the dependence of the shift $\delta_0$, obtained from the solution of Eq. (17), on the positive parameters $p_1$ and $p_2$. At small values of $p_2$, all curves decrease because the contribution of the field acting on the transition from level $\lvert 2 \rangle$ increases (recall that the two-photon detuning is defined as $\delta = \Delta_1 - \Delta_2$). The shift vanishes at $p_1 = p_2$ owing to the symmetrization of the considered double $\Lambda$ scheme. For larger $p_2$ the shift tends to zero for any value of $p_1$. This behavior arises because, at large $p_1$ or $p_2$, the excitation channel through state $\lvert 4 \rangle$ becomes dominant, reducing the system to an effective three-level $\Lambda$ scheme with the excited state $\lvert 4 \rangle$, in which the CPT resonance is known to be free of light shifts \cite{knappe2001characterization}. As a result, between the points $p_2 = p_1$ and $p_2 \to \infty$, an extremum appears where the absolute value of the shift $\delta_0$ reaches a local maximum. It should be emphasized that this feature of the dependence cannot be described within the framework of the conventional dynamic Stark effect.

A significant feature of the shift ${{\delta }_{0}}$ defined by Eq. (17) is the possibility for its dependence on the intensity of the excitation radiation to deviate from the linear law predicted by the dynamic Stark effect framework \cite{affolderbach2005light, yano2014theoretical, Zhu2000}. To analyze this possibility, we introduce the dimensionless parameter $x=\left( \Omega_{1}^{2}+\Omega_{2}^{2} \right)/{{\Gamma }^{2}}$, which, due to the adiabatic approximation used, is small. Here, $x$ is proportional to the sum of the intensities of the two frequency components of the radiation, and thus proportional to the total excitation intensity. We seek the solution of Eq. (17) as a power series expansion in the small parameter $x$:
\begin{equation}
    {{\delta }_{0}}={{\alpha }^{(0)}}+{{\alpha }^{(1)}}x+{{\alpha }^{(2)}}{{x}^{2}}+...,
\end{equation}
where the expansion coefficients ${{\alpha }^{(n)}}$ can be found by successive approximations. Note that these coefficients do not depend on the index $g$, since the coefficients ${{A}_{n}},{{B}_{n}}$ do not depend on it. Moreover, for the solutions of interest, the coefficients satisfy ${{\alpha }^{(0)}}=0$, since the light shift vanishes as the field intensities approach zero.

\begin{figure}
\centering
\includegraphics[width=90mm]{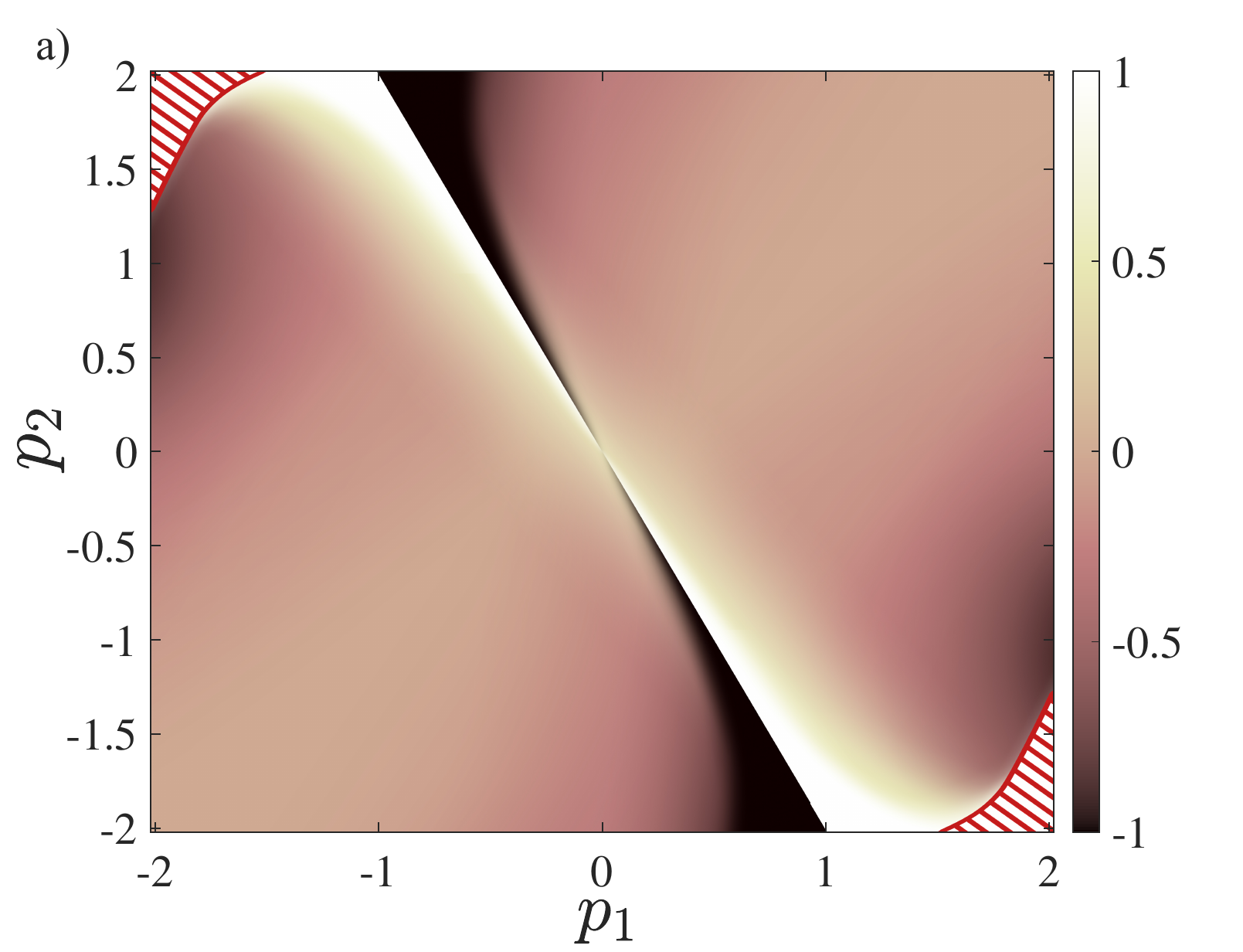}
\includegraphics[width=90mm]{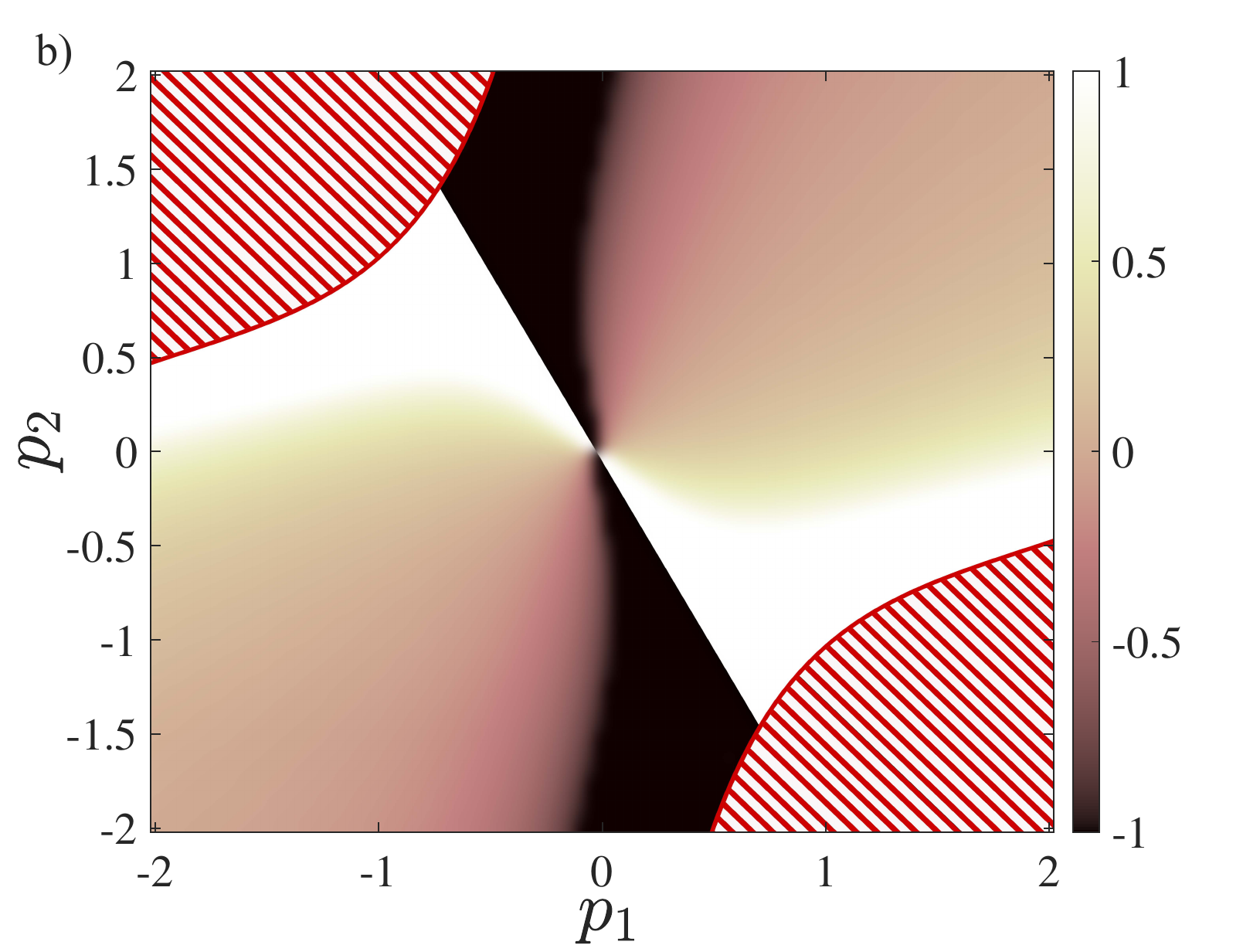}
\caption{Color map of the dependence of ${{\alpha }^{(2)}}/{{\alpha }^{(1)}}$ on ${{p}_{1}}$ and ${{p}_{2}}$ in the range from –1 to 1 along the z-axis for (a) ${{\omega }_{34}}=2\Gamma$ and (b) ${{\omega }_{34}}=0.2\Gamma$ with $\Omega _{1}^{2}=2\Omega _{2}^{2}$. The red hatched regions indicate where ${{\alpha }^{(1)}}$ and ${{\alpha }^{(2)}}$ are undefined.}
\label{color_map}
\end{figure}

We study the dependence of the ratio of the quadratic coefficient to the linear one ${{\alpha }^{(2)}}/{{\alpha }^{(1)}}$ on ${{p}_{1}}$ and ${{p}_{2}}$ using the color maps presented in Fig.~\ref{color_map}. First, note that if ${{p}_{1}}$ and ${{p}_{2}}$ have opposite signs, there exist parameter regions where the CPT resonance disappears, as shown in Fig.~\ref{rho_exc}. In these regions, the concept of the shift (18) loses meaning, and the values of ${{\alpha }^{(2)}}$ and ${{\alpha }^{(1)}}$ become undefined. These areas are marked with red hatching in Fig.~\ref{color_map}. At the same time, in the region of small ${{p}_{1}}$ and ${{p}_{2}}$, the CPT resonance persists even if they differ in sign, since small ${{p}_{1}}$ and ${{p}_{2}}$ imply weak interaction with the fourth level. Furthermore, Fig.~\ref{color_map} shows regions of ${{p}_{1}}$ and ${{p}_{2}}$ where the quadratic coefficient ${{\alpha }^{(2)}}$ becomes larger in absolute value than the linear one ${{\alpha }^{(1)}}$ (dark and light areas in Fig.~\ref{color_map}). The dark regions are of particular interest because there the quadratic contribution subtracts from the linear one, reducing the intensity dependence of the shift. The largest absolute value of the ratio ${{\alpha }^{(2)}}/{{\alpha }^{(1)}}$ occurs near the feature along the line ${{p}_{2}}\Omega_{2}^{2}=-{{p}_{1}}\Omega_{1}^{2}$, where the linear coefficient ${{\alpha }^{(1)}}$ vanishes while ${{\alpha }^{(2)}}$ remains finite. The vanishing of the linear coefficient ${{\alpha }^{(1)}}$ at ${{p}_{2}}\Omega_{2}^{2}=-{{p}_{1}}\Omega_{1}^{2}$ is described by formula (11), which, although losing applicability here, still correctly describes the form of the linear part of the shift’s dependence on ${{p}_{1}}$ and ${{p}_{2}}$. This is explained by the fact that the only possible quadratic dependence of the additional shift ${{\delta }_{D}}$ on ${{p}_{1}}$ and ${{p}_{2}}$, which depends on their signs and changes sign upon swapping indices 1 and 2 (due to the symmetry of the $\Lambda$ scheme), is of the form ${{\delta }_{D}}\sim {{p}_{1}}{{p}_{2}}\left( \Omega_{2}^{2}-\Omega_{1}^{2} \right)$. It is also worth noting that on the line ${{p}_{1}}={{p}_{2}}$, the $\Lambda$ scheme becomes fully symmetric and the light shift disappears, so ${{\alpha }^{(1,2)}}\to 0$. However, the ratio ${{\alpha }^{(2)}}/{{\alpha }^{(1)}}$ tends to a finite value, so the line ${{p}_{1}}={{p}_{2}}$ in Fig.~\ref{color_map} shows no special features. The same holds for the line ${{p}_{1}}=-{{p}_{2}}$ in the case ${{\Omega }_{1}}={{\Omega }_{2}}$, which is not shown in Fig.~\ref{color_map}.

\begin{figure*}
\centering
\includegraphics[width=90mm]{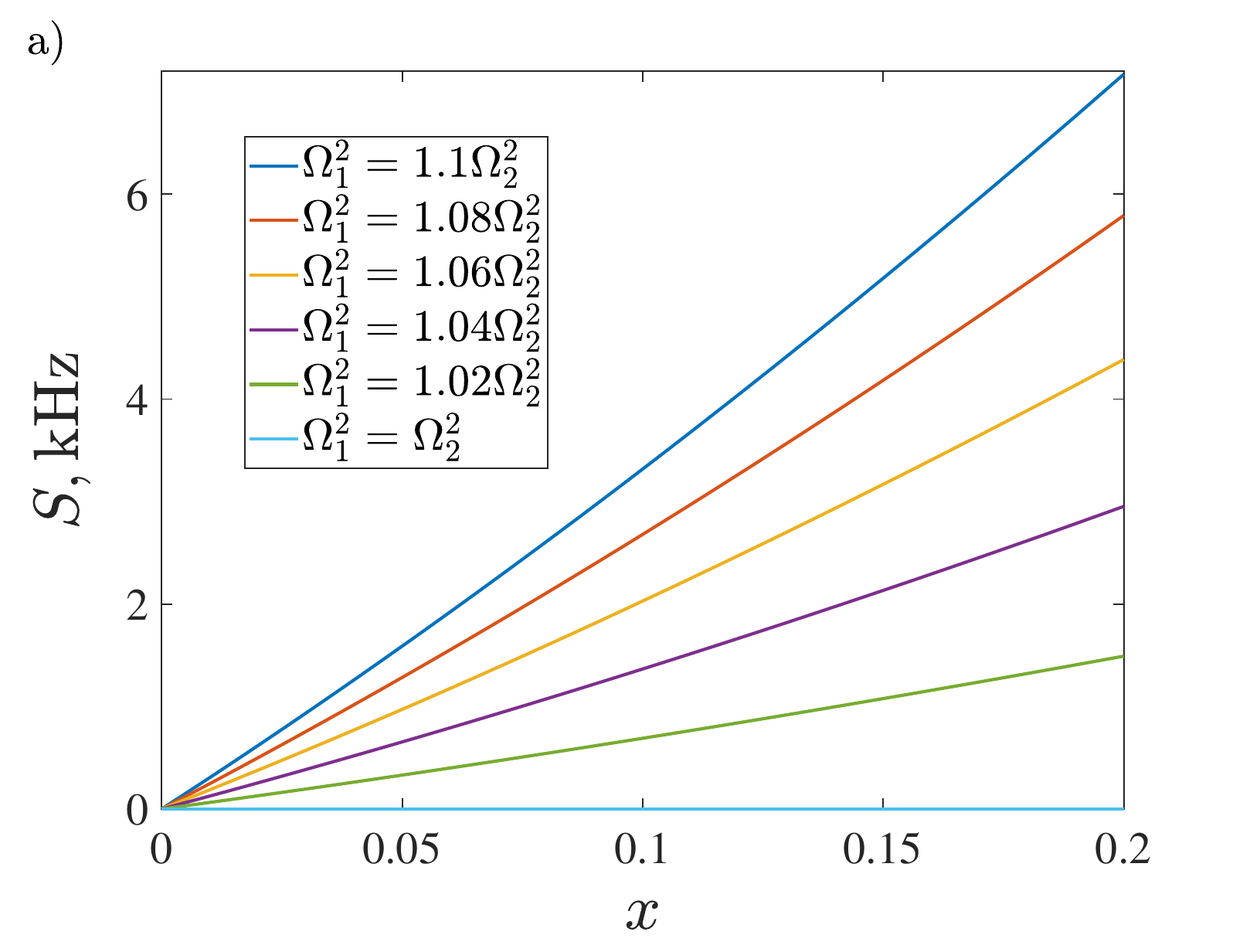}\includegraphics[width=90mm]{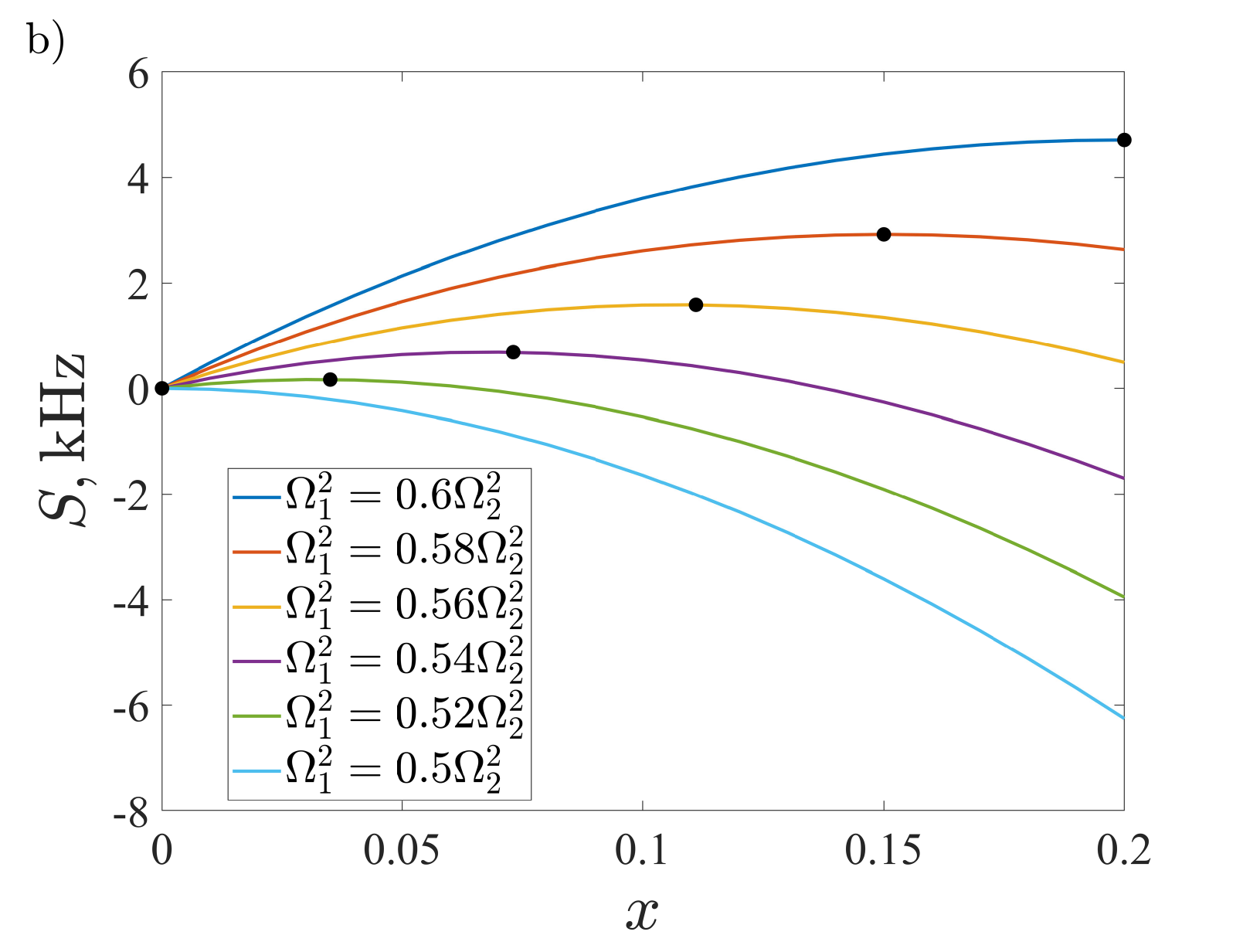}
\caption{Dependences of the CPT resonance shift $S$ on parameter $x$ at (a) ${{p}_{1}}=1$, ${{p}_{2}}=-1$ and (b) ${{p}_{2}}=-0.5$ for ${{\omega }_{34}}=\Gamma$ at various ratios of $\Omega _{1}^{2}$ and $\Omega _{2}^{2}$. Black dots mark the extrema of the dependencies.}
\label{S_x}
\end{figure*}

Figure ~\ref{S_x} illustrates how the dependence of the shift $S={{\delta }_{0}}/2\pi$ on the total intensity (parameter $x$) changes when varying the ratio ${{\Omega }_{1}}/{{\Omega }_{2}}$. The range of variation of the ratio ${{\Omega }_{1}}/{{\Omega }_{2}}$ is chosen to illustrate the shift behavior in the nonlinear region near the line ${{p}_{2}}\Omega_{2}^{2}=-{{p}_{1}}\Omega_{1}^{2}$ in Fig.~\ref{color_map} (dark area). It is seen that in the case ${{p}_{1}}=-{{p}_{2}}=1$ [Fig.~\ref{S_x}(a)], nonlinearity appears weakly and disappears completely at ${{\Omega }_{1}}={{\Omega }_{2}}$, since under these conditions the shift in the considered scheme is absent. For ${{p}_{1}}=1$, ${{p}_{2}}=0.5$ [Fig.~\ref{S_x}(b)], the nonlinearity is more pronounced in the considered range. Approaching the condition ${{p}_{2}}\Omega_{2}^{2}=-{{p}_{1}}\Omega_{1}^{2}$ nullifies the linear contribution and leaves only the nonlinear one. At the same time, the extremum point of the dependence $S(x)$ shifts along the $x$-axis. Thus, by adjusting the ratio ${{\Omega }_{1}}/{{\Omega }_{2}}$, one can minimize the sensitivity of the shift to small changes in the total intensity at arbitrary values.

\section{Conclusion}

In this work, a model has been developed to describe the light shift of the CPT resonance by accounting for the interaction of two radiation components with an atomic system using a $\Lambda$ scheme and takes into account an additional level of excited state. The cases of weak and strong interaction have been analyzed. In the case of weak interaction, an analytical expression for the CPT resonance shift has been obtained, which takes the form of a sum of the Stark shift of the atomic level and an additional shift caused by distortion of the line shape. It is shown that the derived formula remains valid for the observed CPT signals. In the case of strong interaction, it is demonstrated that the dependence of the CPT resonance shift on the total excitation intensity can deviate from the linear law predicted by the dynamic Stark effect under certain relations of the Rabi frequencies of the two excitation components, allowing suppression of the shift’s sensitivity to small intensity variations for arbitrary values.

In conclusion, we note that although these results were obtained for the case of continuous-wave excitation of the CPT resonance, there is reason to believe that the resulting additional shift will also arise when using the Ramsey pulsed method. This is because the specifics of the pulsed excitation method lead to a relation for the shift similar to Eq. (4) in Ref. \cite{pollock2018ac}, which describes a direct proportionality between the shifts of the Ramsey fringes and the steady-state light Stark shift of the CPT resonance. A more detailed analysis of this issue is planned for a future study.
\newpage
\appendix

\section{Analytical solution of the steady-state problem}

Rewrite system (6) for the steady-state case and in the adiabatic approximation:
\allowdisplaybreaks[1]
\begin{eqnarray}
    0=&&-i\left( {{\Omega }_{1}}{{{\tilde{\rho }}}_{13}}-\Omega _{1}^{*}{{{\tilde{\rho }}}_{31}}+{{p}_{1}}{{\Omega }_{1}}{{{\tilde{\rho }}}_{14}}-p_{1}^{*}\Omega _{1}^{*}{{{\tilde{\rho }}}_{41}} \right) \nonumber\\
    &&+{{\gamma }_{31}}{{\rho }_{33}}+{{\gamma }_{41}}{{\rho }_{44}}, \\
    0= &&-i\left( {{\Omega }_{2}}{{{\tilde{\rho }}}_{23}}-\Omega _{2}^{*}{{{\tilde{\rho }}}_{32}}+{{p}_{2}}{{\Omega }_{2}}{{{\tilde{\rho }}}_{24}}-p_{2}^{*}\Omega _{2}^{*}{{{\tilde{\rho }}}_{42}} \right) \nonumber\\
    &&+{{\gamma }_{32}}{{\rho }_{33}}+{{\gamma }_{42}}{{\rho }_{44}}, \\
    0= &&i{{\Omega }_{1}}{{\tilde{\rho }}_{13}}-i\Omega _{1}^{*}{{\tilde{\rho }}_{31}}+i{{\Omega }_{2}}{{\tilde{\rho }}_{23}}-i\Omega _{2}^{*}{{\tilde{\rho }}_{32}} \nonumber\\
    &&-{{\gamma }_{3}}{{\rho }_{33}}, \\
    0=&&i{{p}_{1}}\Omega _{1}^{{}}{{\tilde{\rho }}_{14}}-ip_{1}^{*}\Omega _{1}^{*}{{\tilde{\rho }}_{41}}+i{{p}_{2}}{{\Omega }_{2}}{{\tilde{\rho }}_{24}} \nonumber\\
    &&-ip_{2}^{*}\Omega _{2}^{*}{{\tilde{\rho }}_{42}}-{{\gamma }_{4}}{{\rho }_{44}}, \\
    0=&& -i\Omega _{2}^{*}{{\tilde{\rho }}_{12}}-i\Omega _{1}^{*}{{\rho }_{11}}-{{\delta }_{13}}{{\tilde{\rho }}_{13}}, \\
    0=&& -ip_{2}^{*}\Omega _{2}^{*}{{\tilde{\rho }}_{12}}-ip_{1}^{*}\Omega _{1}^{*}{{\rho }_{11}}-{{\delta }_{14}}{{\tilde{\rho }}_{14}},\\
    0=&& -i\Omega _{1}^{*}{{\tilde{\rho }}_{21}}-i\Omega _{2}^{*}{{\rho }_{22}}-{{\delta }_{23}}{{\tilde{\rho }}_{23}}, \\
    0=&& -ip_{1}^{*}\Omega _{1}^{*}{{\tilde{\rho }}_{21}}-ip_{2}^{*}\Omega _{2}^{*}{{\rho }_{22}}-{{\delta }_{24}}{{\tilde{\rho }}_{24}},\\
    0=&&i\left( \Omega _{1}^{*}{{{\tilde{\rho }}}_{32}}-{{\Omega }_{2}}{{{\tilde{\rho }}}_{13}}+p_{1}^{*}\Omega _{1}^{*}{{{\tilde{\rho }}}_{42}}-{{p}_{2}}{{\Omega }_{2}}{{{\tilde{\rho }}}_{14}} \right) \nonumber\\
    &&-{{\delta }_{12}}{{\tilde{\rho }}_{12}},
\end{eqnarray}
where ${{\delta }_{g3}}=i{{\Delta }_{g}}+{{\Gamma }_{g3}}$, ${{\delta }_{g4}}=i\left( {{\Delta }_{g}}-{{\omega }_{34}} \right)+{{\Gamma }_{g4}}$, ${{\delta }_{12}}=i\left( {{\Delta }_{1}}-{{\Delta }_{2}} \right)+{{\Gamma }_{12}}$.

Expressing optical coherences ${{\tilde{\rho }}_{ge}}$ from (A5)–(A8) and populations ${{\rho }_{ee}}$ from (A3) and (A4), and substituting them into (A9) and (A1), taking into account the approximate normalization ${{\rho }_{11}}+{{\rho }_{22}}\approx 1$, we obtain a system of two equations for ${{\rho }_{11}}$ and ${{\rho }_{12}}$:
\begin{equation}
\begin{aligned}
  & a{{\rho }_{11}}+\operatorname{Re}\left\{ b{{{\tilde{\rho }}}_{12}} \right\}=f, \\ 
 & c{{\rho }_{11}}+d{{{\tilde{\rho }}}_{12}}=h, 
\end{aligned}
\end{equation}
where
\allowdisplaybreaks[1]
\begin{align}
  a=& -\operatorname{Re}\left\{ \frac{{{\left| {{\Omega }_{1}} \right|}^{2}}}{{{\gamma }_{31}}{{\delta }_{13}}}\left( 1+{{\left| {{p}_{1}} \right|}^{2}}\left( 1-\Delta {{\gamma }_{1}} \right)\frac{{{\delta }_{13}}}{{{\delta }_{14}}} \right)\right. \nonumber \\
  &\left. + \frac{{{\left| {{\Omega }_{2}} \right|}^{2}}}{{{\gamma }_{32}}{{\delta }_{23}}}\left( 1+{{\left| {{p}_{2}} \right|}^{2}}\left( 1-\Delta {{\gamma }_{2}} \right)\frac{{{\delta }_{23}}}{{{\delta }_{24}}} \right) \right\}, \nonumber \\ 
  b=&-\frac{{{\Omega }_{1}}\Omega _{2}^{*}}{{{\gamma }_{31}}{{\delta }_{13}}}\left( 1+{{p}_{1}}p_{2}^{*}\left( 1-\Delta {{\gamma }_{1}} \right)\frac{{{\delta }_{13}}}{{{\delta }_{14}}} \right) \nonumber \\
 &+ \frac{{{\Omega }_{1}}\Omega _{2}^{*}}{{{\gamma }_{32}}\delta _{23}^{*}}\left( 1+{{p}_{1}}p_{2}^{*}\left( 1-\Delta {{\gamma }_{2}} \right)\frac{\delta _{23}^{*}}{\delta _{24}^{*}} \right), \nonumber \\ 
  c=&\frac{\Omega _{1}^{*}{{\Omega }_{2}}}{\delta _{23}^{*}}\left( 1+{{p}_{2}}p_{1}^{*}\frac{\delta _{23}^{*}}{\delta _{24}^{*}} \right) \nonumber \\
  &- \frac{\Omega _{1}^{*}{{\Omega }_{2}}}{{{\delta }_{13}}}\left( 1+{{p}_{2}}p_{1}^{*}\frac{{{\delta }_{13}}}{{{\delta }_{14}}} \right), \nonumber \\ 
  d=&-\frac{{{\left| {{\Omega }_{1}} \right|}^{2}}}{\delta _{23}^{*}}\left( 1+{{\left| {{p}_{1}} \right|}^{2}}\frac{\delta _{23}^{*}}{\delta _{24}^{*}} \right) \nonumber \\
  &- \frac{{{\left| \Omega _{2}^{*} \right|}^{2}}}{{{\delta }_{13}}}\left( 1+{{\left| {{p}_{2}} \right|}^{2}}\frac{{{\delta }_{13}}}{{{\delta }_{14}}} \right)-{{\delta }_{12}}, \nonumber \\ 
  f=&-\frac{{{\left| {{\Omega }_{2}} \right|}^{2}}}{{{\gamma }_{32}}}\operatorname{Re}\left\{ \frac{1}{{{\delta }_{23}}}\left( 1+{{\left| {{p}_{2}} \right|}^{2}}\left( 1-\Delta {{\gamma }_{2}} \right)\frac{{{\delta }_{23}}}{{{\delta }_{24}}} \right) \right\}, \nonumber \\ 
  h=&\frac{\Omega _{1}^{*}{{\Omega }_{2}}}{\delta _{23}^{*}}\left( 1+{{p}_{2}}p_{1}^{*}\frac{\delta _{23}^{*}}{\delta _{24}^{*}} \right).
\end{align}

Here, the notation \[\Delta {{\gamma }_{g}}=\left( \frac{{{\gamma }_{4g}}}{{{\gamma }_{3g}}}-\frac{{{\gamma }_{4{g}'}}}{{{\gamma }_{3{g}'}}} \right)\frac{{{\gamma }_{3g}}}{{{\gamma }_{4}}} \] is used.

The solution of system (A10) taking into account the complex nature of ${{\tilde{\rho }}_{12}}$ is given by
\begin{equation}
    \begin{aligned}
  & {{\rho }_{11}}=\frac{-{{\left| d \right|}^{2}}f+b{{d}^{*}}h+d{{b}^{*}}{{h}^{*}}}{-a{{\left| d \right|}^{2}}+d{{b}^{*}}{{c}^{*}}+bc{{d}^{*}}}, \\ 
 & {{{\tilde{\rho }}}_{12}}=\frac{c{{d}^{*}}f+{{b}^{*}}{{c}^{*}}h-a{{d}^{*}}h-c{{b}^{*}}{{h}^{*}}}{-a{{\left| d \right|}^{2}}+d{{b}^{*}}{{c}^{*}}+bc{{d}^{*}}}. \\ 
\end{aligned}
\end{equation}

Expressions for the coefficients (A11) can be simplified in the uniform decay model (7) (where $\Delta {{\gamma }_{g}}=0$), using the weak interaction approximation: ${{\omega }_{34}}\gg \Gamma $. In the region of small detunings ${{\Delta }_{g}}\ll \Gamma $, it is possible to approximate ${{\delta }_{g3}}\approx \Gamma $ and ${{\delta }_{g4}}\approx -i{{\omega }_{34}}+\Gamma$, assuming the absorption contour is sufficiently broad. As a result, neglecting terms of second order smallness relative to $\Gamma /{{\omega }_{34}}$, we obtain
\begin{equation}
\begin{aligned}
  & a\approx -\frac{4}{\Gamma \gamma }\left( {{\left| {{\Omega }_{1}} \right|}^{2}}+{{\left| {{\Omega }_{2}} \right|}^{2}} \right), \\ 
 & b\approx -i\frac{4{{p}_{1}}p_{2}^{*}{{\Omega }_{1}}\Omega _{2}^{*}}{\Gamma \gamma }G, \\ 
 & c\approx -\frac{2ip_{1}^{*}{{p}_{2}}\Omega _{1}^{*}{{\Omega }_{2}}}{\Gamma }G, \\ 
 & d\approx -{{\gamma }_{D}}-i\left( \delta -{{\delta }_{AC}} \right), \\ 
 & f\approx -\frac{4{{\left| {{\Omega }_{2}} \right|}^{2}}}{\Gamma \gamma }, \\ 
 & h\approx \frac{\Omega _{1}^{*}{{\Omega }_{2}}}{\Gamma }\left( 1+p_{1}^{*}{{p}_{2}}iG \right).
\end{aligned}
\end{equation}

Substituting (A13) into (A12) and keeping terms of first order smallness relative to $G$, we finally obtain expressions (8), (9).

\section{Coefficients in the expression for the imaginary part of susceptibility}

Express the optical coherences ${{\tilde{\rho }}_{ge}}$ from (A.5)–(A.8):
\begin{equation}
\begin{aligned}
        &{{\tilde{\rho }}_{g3}}=\left( -i\Omega _{{{g}'}}^{*}{{\rho }_{g{g}'}}-i\Omega _{g}^{*}{{\rho }_{gg}} \right)/{{\delta }_{g3}},\\
        &{{\rho }_{g4}}= \left( -ip_{{{g}'}}^{*}\Omega _{{{g}'}}^{*}{{\rho }_{g{g}'}}-ip_{g}^{*}\Omega _{g}^{*}{{\rho }_{gg}} \right)/{{\delta }_{g4}},
\end{aligned}
\end{equation}

We limit ourselves to the uniform decay model (7) and the case of real ${{\Omega }_{g}}$ and ${{p}_{g}}$. Substituting the solutions (A12) into (B1), we find the susceptibilities from (13) in the form (16), where

\allowdisplaybreaks[1]

\begin{widetext}
\begin{align*}
A_1 &= -64(p_1 - p_2)^2\left\{\Gamma^2(\tilde{p}_1^2\Omega_1^2 + \tilde{p}_2^2\Omega_2^2)^2 
     + [\tilde{p}_1^2\Omega_1^4 + 2(1 + p_1p_2)\Omega_1^2\Omega_2^2 + \tilde{p}_2^2\Omega_2^4]\omega_{34}^2\right\},\\[4pt]
A_2 &= 64(p_1 - p_2)\omega_{34} \left\{ -p_2\Omega_1^4 - p_2^3\Omega_1^2\Omega_2^2 + p_1^3\Omega_1^2(2\Gamma^2 + \Omega_1^2 - \Omega_2^2)\right.\\
&\left.\qquad\qquad + p_2\Omega_2^2(\Gamma^2 + 2\omega_{34}^2 + \Omega_2^2) 
     + p_1[2\Gamma^2\Omega_1^2 + \Omega_1^4 - p_2^2\Omega_1^2 - \tilde{p}_2^2\Omega_2^4 + 2\tilde{p}_2^2\Omega_1^2\Omega_2^2]\right\},\\[4pt]
A_3 &= -16\left\{4\Gamma^2\Omega_1^2\Omega_2^2 + 4\omega_{34}^4 - 4p_2^2\omega_{34}^2\Omega_1^2 + p_2^6\Omega_1^4 - 2p_1^5p_2\Omega_1^4 + p_2^2\Omega_1^4\right.\\
&\qquad\qquad + 4p_2^2\omega_{34}^2\Omega_2^2 + 2p_2^2\Omega_1^2\Omega_2^2 + 2p_2^4\Omega_1^2\Omega_2^2 + p_2^2\Omega_2^4 + 2p_2^4\Omega_2^4 + p_2^6\Omega_2^4\\
&\qquad\qquad - 4p_1^3p_2(\Omega_1^2 + \tilde{p}_2^2\Omega_2^2)\Omega_1^2 
               - 2p_1p_2(\Omega_1^2 + \tilde{p}_2^2\Omega_2^2)^2 
               + p_1^4(\Omega_1^4 + 2\tilde{p}_2^2\Omega_1^2\Omega_2^2)\\
&\qquad\qquad -4\omega_{34}^2(-2\omega_{34}^2 + p_1^4\Omega_1^2 - 2p_1^3p_2\Omega_1^2 + p_2^4\Omega_2^2 + p_2^2[-\omega_{34}^2 + \Omega_1^2 + \Omega_2^2]\\
&\qquad\qquad - 2p_1p_2(\Omega_1^2 + \tilde{p}_2^2\Omega_2^2) + p_1^2[-\omega_{34}^2 + \tilde{p}_2^2\Omega_1^2 + \tilde{p}_2^2\Omega_2^2])\\
&\left.\qquad\qquad + p_1^2[(1 + 2p_2^2)\Omega_1^4 - 4\omega_{34}^2\Omega_2^2 + \tilde{p}_2^2\Omega_2^4 + 2\Omega_1^2(2\omega_{34}^2 + \tilde{p}_2^2\Omega_2^2)]\right\},\\[4pt]
A_4 &= 32\omega_{34}(p_1 - p_2)\left\{p_1^3\Omega_1^2 - p_1\left[\Gamma^2 + \Omega_1^2 - 2\Omega_2^2\right]
      + p_2\left[-\Gamma^2 + 2\Omega_1^2 + (-1 + p_2^2)\Omega_2^2\right]\right\},\\[4pt]
A_5 &= 16\bigl[-2\Gamma^2\tilde{p}_1^2\tilde{p}_2^2 + 2\omega_{34}^2 - p_2^2\omega_{34}^2 + p_1^4\Omega_1^2 
      - 2p_1^3p_2\Omega_1^2 + p_2^2\Omega_1^2 + p_2^2\Omega_2^2 + p_2^4\Omega_2^2\\
&\qquad\qquad - 2p_1p_2(\Omega_1^2 + \tilde{p}_2^2\Omega_2^2) + p_1^2(-\omega_{34}^2 + \tilde{p}_2^2\Omega_1^2 + \tilde{p}_2^2\Omega_2^2)\bigr],\\[4pt]
A_6 &= 16\omega_{34}(p_2^2 - p_1^2),\\
A_7 &= -4\tilde{p}_1^2\tilde{p}_2^2,\\
B_1 &= -64\omega_{34}^4(\Omega_1^2 + \Omega_2^2)^3 - 64\Gamma^4(\tilde{p}_1^2\Omega_1^2 + \tilde{p}_2^2\Omega_2^2)^3 - 64\Gamma^2\omega_{34}^2(\Omega_1^2 + \Omega_2^2)\times\left\{(2 + 3p_1^2 + p_1^4)\Omega_1^4\right.\\
&\left.\qquad\qquad + \left[4 + 3p_2^2 + p_1^2(3 + 2p_2^2)\right]\Omega_1^2\Omega_2^2 + (2 + 3p_2^2 + p_2^4)\Omega_2^4\right\},\\[4pt]
B_2 &= 64\omega_{34}[\tilde{p}_1^2\Omega_1^2-\tilde{p}_2^2\Omega_2^2]\left[2\Gamma^2\omega_{34}^2(\Omega_1^2 + \Omega_2^2) - \omega_{34}^2(\Omega_1^2 + \Omega_2^2)^2 + 2\Gamma^4(\tilde{p}_1^2\Omega_1^2 + \tilde{p}_2^2\Omega_2^2)\right],\\[4pt]
B_3 &= -64\Gamma^2(\tilde{p}_1^2\Omega_1^2 + \tilde{p}_2^2\Omega_2^2) + 64\Gamma^4\left\{\tilde{p}_1^2\Omega_1^4 - (2 + p_2^2)\omega_{34}^2\Omega_2^2 + \tilde{p}_2^2\Omega_2^4 \right.\\
&\left.\qquad + \Omega_1^2[-(\,2 + p_1^2)\omega_{34}^2 + 2\tilde{p}_1^2\tilde{p}_2^2\Omega_2^2]\right\} - 16\omega_{34}^2(\Omega_1^2 + \Omega_2^2)\times\left\{(p_1^4 - 2 - p_1^2)\Omega_1^4\right.\\
&\qquad \left. + (p_2^4 - 2 - p_2^2)\Omega_2^4 - 4\omega_{34}^2\Omega_2^2 - 4\omega_{34}^2\Omega_1^2 + [-4 - p_2^2 + p_1^2(-1 + 2p_2^2)]\Omega_2^2\right\}\\
&\qquad - 32\Gamma^2[\tilde{p}_1^2\Omega_1^6 + 2\omega_{34}^4\Omega_2^2 - 4\omega_{34}^2\Omega_2^4 + \tilde{p}_2^2\Omega_2^6\\
&\qquad + \Omega_1^4(-4\omega_{34}^2 + 3\tilde{p}_1^2\tilde{p}_2^2\Omega_2^2) + \Omega_1^2(2\omega_{34}^4 - 8\omega_{34}^2\Omega_2^2 + 3\tilde{p}_1^2\tilde{p}_2^2\Omega_2^4)],\\[4pt]
B_4 &= 32\omega_{34}(p_1^2\Omega_1^2 - p_2^2\Omega_2^2)[-2\Gamma^4 + \omega_{34}^2(\Omega_1^2 + \Omega_2^2) + 2\Gamma^2(\tilde{p}_1^2\Omega_1^2 + \tilde{p}_2^2\Omega_2^2)],\\[4pt]
B_5 &= -4\left\{\tilde{p}_1^2\Omega_1^6 + 4\omega_{34}^4\Omega_2^2 + 8\omega_{34}^2\Omega_2^4 + \Omega_2^6 + 3p_2^2\Omega_2^6 + 3p_2^4\Omega_2^6 + p_2^6\Omega_2^6\right.\\
&\qquad + 12\Gamma^2(\tilde{p}_1^2\Omega_1^2 + \tilde{p}_2^2\Omega_2^2) + \Omega_1^4(8\omega_{34}^2 + 3\tilde{p}_1^2\tilde{p}_2^2\Omega_2^2)\\
&\qquad + \Omega_1^2(4\omega_{34}^4 + 16\omega_{34}^2\Omega_2^2 + 3\tilde{p}_1^2\tilde{p}_2^2\Omega_2^4)\\
&\qquad \left.- 8\Gamma^2\left[\tilde{p}_1^2\Omega_1^4 - p_2^2\omega_{34}^2\Omega_2^2 + \tilde{p}_2^2\Omega_2^4 + \Omega_1^2(-p_1^2\omega_{34}^2 + 2\tilde{p}_1^2\tilde{p}_2^2\Omega_2^2)\right]\right\},\\[4pt]
B_6 &= 8\omega_{34}(p_1^2\Omega_1^2 - p_2^2\Omega_2^2)([)-4\Gamma^2 + \tilde{p}_1^2\Omega_1^2 + \tilde{p}_2^2\Omega_2^2),\\[4pt]
B_7 &= 4\left\{\tilde{p}_1^2\Omega_1^4 - (p_2^2 - 2)\omega_{34}^2\Omega_2^2 + \tilde{p}_2^2\Omega_2^4 - 3\Gamma^2(\tilde{p}_1^2\Omega_1^2 + \tilde{p}_2^2\Omega_2^2)\right.\\
&\qquad \left.+ \Omega_1^2[-(p_1^2 - 2)\omega_{34}^2 + 2\tilde{p}_1^2\tilde{p}_2^2\Omega_2^2]\right\},\\[4pt]
B_8 &= -4\omega_{34}(p_1^2\Omega_1^2 - p_2^2\Omega_2^2),\\
B_9 &= -(\tilde{p}_1^2\Omega_1^2 + \tilde{p}_2^2\Omega_2^2),
\end{align*}
where $1+p_g^2 \equiv \tilde{p}_g^2$.
\end{widetext}

\nocite{*}

\providecommand{\noopsort}[1]{}\providecommand{\singleletter}[1]{#1}%

\end{document}